\documentclass{llncs}
\usepackage{graphicx}
\usepackage{float}
\usepackage{subfigure}
\usepackage[hidelinks]{hyperref}
\usepackage{color}
\usepackage{cite}

\usepackage[prependcaption, textsize=scriptsize, linecolor=red, backgroundcolor=red!25, bordercolor=red, textwidth=3cm]{todonotes}

\begin{document}
\title{Dynamic community detection into analyzing of wildfires events}
%
%
\author{Alessandra M. M. M. Gouvêa\inst{1}\orcidID{0000-0002-0512-0162} \and
Didier A. Vega-Oliveros\inst{2,3}\orcidID{0000-0001-9569-3775} \and
Moshé Cotacallapa\inst{2}\orcidID{0000-0002-1949-2696} \and
Leonardo N. Ferreira\inst{2}\orcidID{0000-0003-1445-0590} \and
Elbert E. N. Macau\inst{1,2}\orcidID{0000-0002-6337-8081} \and
Marcos G. Quiles\inst{1}\orcidID{0000-0001-8147-554X}
}
\authorrunning{Gouvêa et al.}
%
\institute{Federal University of São Paulo, São José dos Campos, SP, Brazil \and
National Institute for Space Research,  São José dos Campos, SP, Brazil \and Institute of Computing, University of Campinas, Campinas, SP, Brazil \\
\email{\{alessandra.marli,elbert.macau,quiles\}@unifesp.br  \{didier.oliveros,frank.moshe,leonardo.ferreira\}@inpe.br}}

\maketitle              
\begin{abstract}
The study and comprehension of complex systems are crucial intellectual and scientific challenges of the 21st century. In this scenario, network science has emerged as a mathematical tool to support the study of such systems. Examples include environmental processes such as wildfires, which are known for their considerable impact on human life. However, there is a considerable lack of studies of wildfire from a network science perspective. Here, employing the chronological network concept --- a temporal network where nodes are linked if two consecutive events occur between them--- we investigate the information that dynamic community structures reveal about the wildfires' dynamics. Particularly, we explore a two-phase dynamic community detection approach, i.e., we applied the Louvain algorithm on a series of snapshots. Then we used the Jaccard similarity coefficient to match communities across adjacent snapshots. Experiments with the MODIS dataset of fire events in the Amazon basing were conducted. Our results show that the dynamic communities can reveal wildfire patterns observed throughout the year.

\keywords{Community Detection \and Temporal Networks \and Wildfire \and Fire Activity \and Geographical Data Modeling}
\end{abstract}
\section{Introduction}

Wildfires are described as any uncontrolled fire in combustible vegetation that occurs in the countryside or wilderness areas~\cite{diaz2016modeling}. Such environmental process has a significant impact on human life \cite{vega2019spatio}. For example, it is responsible for damage to properties \cite{dey2018review}; it may affect the forest dynamics, potentially changing hydrological processes and cloud microphysics -- due to the aerosols emissions of gases and particles, which results in a considerable impact on climate \cite{mishra2015co}. Besides, as a source of CO$_2$ and particulate matter, it contributes to the greenhouse and global warming \cite{diaz2016modeling} effects. Therefore, modeling this phenomenon, especially regarding its spatial incidence and extension, is a relevant task to support governments and public agencies in the control and risk management of wildfire seasons. Also, it is essential to global warming and climate change researches to understand the related facts to spatial incidence and the size of burned areas \cite{diaz2016modeling}. 

Fire data are usually available in the form of spatio-temporal events, i.e., given a geographical area of study, some measurements of interest are collected in a specific time \cite{ferreira2020}. One common approach to model spatio-temporal datasets is the functional network, which consists of dividing the geographical area into grid cells that represent the nodes, and the edges hold the similarity between pairs of collected measures in the form of time-series. Formally, the network $G=(V,E)$ is formed by the set $V$ of nodes and a set $E$ of edges, where $v_i \in V$ is a cell grid and $(v_i,v_j) \in E$ is the edge defined by a process which consists of computing and linking the vertices if the correlation coefficients between the underlying time-series of spatial grid $v_i$ and $v_j$ are higher than a given threshold. Modeling spatio-temporal data by functional networks, also known as correlation networks, have been successfully used as tools to represent complex systems in a wide range of domains, in which were notably evidenced their potential to identify valuable information \cite{fan2017network,meng2018forecasting,tsonis2008topology,zemp2017deforestation,zhou2015teleconnection}. For instance, analyzing global climate teleconnections \cite{zhou2015teleconnection}, predictions of El-Ni\~no and exploring its impacts around the world \cite{fan2017network,meng2018forecasting,tsonis2008topology}. 

Although network science has a set of well-established tools to model complex systems, few works have modeled wildfires through networks. Moreover, some challenges and issues need attention when using correlation networks. For example, we need to find the ideal correlation threshold to connect nodes and define the time series length to find statistically significant correlations. In this way, it is not possible to elucidate, without trial and error, if the defined construction process can capture the expected temporal patterns. Additionally, for real-world spatio-temporal event data, the correlation-based networks may not be applicable; that happens because only short-length time series are available or the time-series are very sparse with a large number of zeros, which difficulties to evaluate how much the system changed between short time intervals. As an alternative to the correlation networks, some initial works have explored a different approach to construct the network, connecting the nodes by the chronological order of occurrence of the events~\cite{abe2006complex,ferreira2018towards,vega2019spatio}, also know as chronological networks or Chronnets~\cite{vega2019spatio,ferreira2020B}. However, the previous works disregarded the temporal community detection in terms of its spatial incidence and extension.

In this work, we aim to analyze the temporal information stored on the wildfire dataset by using temporal chronnets. Our analysis differs from previous works~\cite{vega2019spatio,Gao2020,ferreira2020B} once we seek to validate the results of the temporal community detection methods when modeling the wildfires, mainly concerning its spatial incidence and extension. Our methodology can reveal where and how often specific fire event patterns occur over the years. We also provide a temporal-based model to wildfires in the Amazon basin, which has significant contributions to researches and agencies concerned with illegal deforestation and climate change since the Amazon basin is responsible for, on average, 15\% of global fire emissions per year \cite{van2010global}.

The paper is organized as follow: Section \ref{sec_RelatedWork} describes the related works with special attention on methodology and obtained results by Vega-Oliveros et. al. \cite{vega2019spatio,ferreira2020B}; Section \ref{sec:materialAndMethods} describes the data set and presents some of the concepts related to community detection and temporal networks; Section \ref{sec_methodology} explains the proposed methodology together with all the setup to conduct our experiments; Section \ref{sec_results}  highlights the obtained results and, finally, Section \ref{sec_conFW} presents the conclusion and points out some future works.

\section{Related Work}
\label{sec_RelatedWork}

Notwithstanding forest ecosystems are prime examples of complex systems \cite{perry1994forest}, network science has been rarely applied in forest ecology and management~\cite{filotas2014viewing}. 
In the case of wildfires, such reality has not seemed to be changed over the years, even though network science is a well-established framework to support studies aiming to model and understand complex systems. We performed a simple search on the Scopus database\footnote{\url{https://www.scopus.com} -- one of the largest and broadly databases indexing peer-reviewed scientific articles and citations} looking for works indexed by the query ``wildfires'' AND ``complex network'' OR ``network science'', which shows the few published works related to this topic, returning less than twenty articles. However, among the results, only the work of Vega-Oliveros et al. \cite{vega2019spatio,ferreira2020B} proposed to represent wildfires data by networks. The remaining returned works were related to off topics, like fire networks for emergency and management, evacuation models, and others.

In \cite{vega2019spatio}, the authors analyzed wildfire data from the Moderate Resolution Imaging Spectroradiometer (MODIS) operated by the National Aeronautics and Space Administration (NASA), showing the applicability of the proposed Spatio-temporal network model. They constructed a dataset of Chronological temporal networks of 15 years of wildfire events from the Amazon basin, which is publicly available (for more details, please see Section~\ref{sec:dataset}). With the constructed set of snapshots, the authors calculated the normalized entropy for each temporal network based on the frequency of fire events, in which higher the entropy, higher the fire activity, and more heterogeneous the network~\cite{vega2019spatio}. The entropy distribution showed a pattern on fire season starting at the beginning of winter and finishes in the middle of summer, which corresponds to the predominant dry season on the Amazon basin. The authors also verified how the nodes are related over time, given the temporal networks. They defined a centrality-series similarity network where they detected twelve community structures by the modular community detection method \cite{lambiotte2014random}. Although they incorporated the temporal information to detected communities from the temporal network representation of wildfire events, the analysis was performed from a static perspective. Additionally, even that the construction process disregarded the geographical locations of nodes, the detected communities seem to respect some geographical order.

Recently, Gao et al. \cite{Gao2020}  proposed a method for mining temporal networks by representative nodes in a community identification approach. They tackled the problem from the change detection perspective, where each stable temporal state of the dataset is represented as a community. In this way, the method has a low computational cost and is applicable for data stream mining. The authors showed the effectivity of the proposed method in some artificial datasets, and in a case study analyzing wildfire events from the same temporal Chronnets dataset~\cite{vega2019spatio,ferreira2020B}. As a result, they also detected two central communities in the Amazon region, each one corresponding to different periods of the year: the south-hemisphere winter season, with a high tendency of fires, and the south-hemisphere summer season, with a low frequency of fires~\cite{Gao2020}. However, these communities represent the global state of the wildfire system in the Amazon basin, not the micro spatial-temporal particularities or patterns into the microregions.

This work differs from the studies mentioned above in many aspects: whereas the authors detected a pattern on fire season that starts at the beginning of winter and finishes in the middle of summer, here, we aim to determine how this pattern evolves, which geographical regions are involved, and what are their spatial incidence and extension over time. We used the same configuration setup reported in \cite{vega2019spatio}; however, we performed our analysis from a dynamic perspective and including the geolocalization distance between the nodes. We verified the use of dynamic community detection processes to model wildfires data regarding the spatial incidence and extension over time. The before is because the processes used to detect such structures are claimed to be an unsupervised model that may help determine the mechanisms governing the network evolution as a whole \cite{aggarwal2014evolutionary}. Therefore, this work is an endeavor to answer: (i) is it possible to model fire dynamics through community detection analysis? (ii)  is it possible to find fire patterns in a given geographical area? Moreover, (iii) do exist outliers on fire dynamics patterns over the years?.

\section{Material and Methods}\label{sec:materialAndMethods}

\subsection{Dataset}
\label{sec:dataset}

It is noteworthy that on a global scale, the fire activity is collected through satellite instruments. We used the dataset, public available, of network snapshots of wildfire data\footnote{\url{https://github.com/fire-networks}} from Vega-Oliveros et al.~\cite{vega2019spatio}. The wildfire data is from the last version (C6) of Moderate Resolution Imaging Spectroradiometer (MODIS), a satellite instrument presented in Aqua and Terra satellites operated by the National Aeronautics and Space Administration (NASA). This dataset encompasses fifteen years (from 2003 to 2018) of fire events from a region of the Amazon basin, located between longitude $70^{\circ}W$, $50^{\circ}W$ and latitude $15^{\circ}S$, $5^{\circ}N$. The selected fire events are those with detection confidence above 70\%, where each register is formed by the fields (i) UTC date and hour in which the fire event was captured by the satellite and  (ii) the geographic coordinates of the event. We highlight that all the events have a different timestamp in this dataset, even though events from different geographic areas may simultaneously occur. The reason is that the satellites sequentially scan the earth's surface so that each event is captured at a time.  

The construction process consists of three steps~\cite{vega2019spatio}: (i) \textbf{Grid Division}, where the geographic region in the study is divided into grid cells, and a node on the network represents each cell. (ii) \textbf{Time Length}, which allows the slicing of the network into specific periods, e.g., the whole time, fixed or dynamic intervals. (iii) \textbf{Links Construction}, which consists of creating an edge between the grid cells where two successive events have occurred, disregarding the geographical distance between the nodes. The authors performed a sensibility analysis to set the parameters, where they concluded that an optimal grid-division for the case of the Amazon basin region is $30x30$, and temporal division in periods of seven days. The dataset is then composed of a set of snapshots, $G = {G_0, G_1, ..., G_l}$, in consecutive and not overlapping intervals of $\Delta_{t}=7$ days, beginning from Jan. 01, 2003 to Jan. 24, 2018. $G_0$ is formed by the events that occurred between Jan. 01, 2003 to Jan. 07, 2003, and $G_1$ for all events from Jan. 08, 2003 to Jan. 14, 2003, and suchlike, totalizing 786 snapshots.

\subsection{Complex Networks}
\label{sec_CommDectTN}

Most of the phenomena that arouse scientific interest can be described through a set of components that are connected in some way \cite{newman2003structure}. The strategy of describing phenomena focusing only on its fundamental units is the base of the network theory. Formally, a network is a mathematical abstraction, $\mathcal{G} = \{\mathcal{V},\,\mathcal{E}\}$, where $\mathcal{V}$ is a set of vertices representing phenomenon components, $\mathcal{E}$ is the set of edges which through a function $\psi$ describe the relationship between the components. Network science assumed that the topological structures --- i.e., the tangled set of vertices and edges --- hold information that can provide means to analyze and guide the understanding of the represented system. Therefore, the studies consist of obtaining a set of structural measures calculated from the network. Among such measures, the communities are one of the essential topological structures in real networks \cite{fortunato2010community}.

Community structure can be intuitively described as being a set of vertices more likely to share edges than vertices from other parts of the graph, where this greater rate of shared edges is explained by the similarity between the vertices which indicates the sharing of a common property that differs from the other vertices of the graph. However, although this intuitive description is well accepted in literature, it is noteworthy that there is no single formal definition that guides the process of identifying communities \cite{rossetti2018community}. Detecting these structures is important in several aspects ranging from practical applications to the study of networks as entities that evolve. Regarding network evolution over time, vertices organization into communities can be seen as an unsupervised model of the entire network \cite{aggarwal2014evolutionary}, so that variations on communities life cycle, like its detection (birth), split, merge, expansion, contraction, and inability to find it in a given time (i.e., death) can provide useful information regarding the overall evolution of the network \cite{gupta2011evolutionary}. The studies about the community life cycle are held by temporal network literature.

\subsection{Temporal Networks}

We are especially interested in modeling and understanding the wildfires' dynamics over time, mainly involving the spatial incidence and extension. In this scenario, the set of vertices and edges receive a new degree of freedom: the time. There are various ways to incorporate time information into a graph, and Holme~\cite{holme2015modern} made an adequate description. Two main approaches to deal with the temporal representation of networks are adopted. The first is the set of techniques that allow a significant loss of temporal information: such techniques aims to obtain static networks through some mechanism that captures both the temporal and topological properties, allowing the loss of information during the process, in an attempt to use the literature on static networks \cite{holme2013temporal}. The second approach is devoted to scenarios where such information is of great value for understanding the phenomenon under study. In this scenario, we have the techniques that need to deal with the time representation losing as little temporal information as possible. 

The representation by snapshots, i.e., a sequence of static graphs, are often used given the balance between the complexity of the analysis and the loss of temporal information \cite{rossetti2018community}. Snapshots allow the use of the static network literature in each snapshot; moreover, such representation does not require considerable additional efforts in the study of temporal aspects. As mentioned in Section \ref{sec:dataset}, we have conducted our research considering the snapshots dataset public available in \cite{vega2019spatio}.

A comprehensive set of methods to extract community structures from snapshots and other kinds of temporal network representation is described by Rossetti et al. \cite{rossetti2018community}. In this work, we use Louvain \cite{blondel2008fast}, a classical and well-known algorithm for community detection in the static network. To deal with the aspects of temporal variation on community structure, we followed the methodology described by Sun et al. \cite{sun2015matrix}, which describes how to track communities aiming to obtain their life cycle from a set of independent communities extracted through distinct snapshots. Other techniques published in the Literature to track communities in an attempt to describe their life-cycle is presented by Dakiche et al. \cite{dakiche2019tracking}.  Finally, it is noteworthy that the methodology described by Sun et al. \cite{sun2015matrix} does not include the concept of a resurgence, i.e., a community which is detected in $t_i$ and it is not observed for a while until it occurs again in $t_{i+{\Delta}t}$. The concept of resurgence is crucial to our analysis so that we can use it to signal the temporal periodicity of a community. We solve this situation by employing the Jaccard similarity coefficient to search for these communities.

\section{Methodology}
\label{sec_methodology}

\begin{figure}[H]
  \centering
	\includegraphics[scale=0.13]{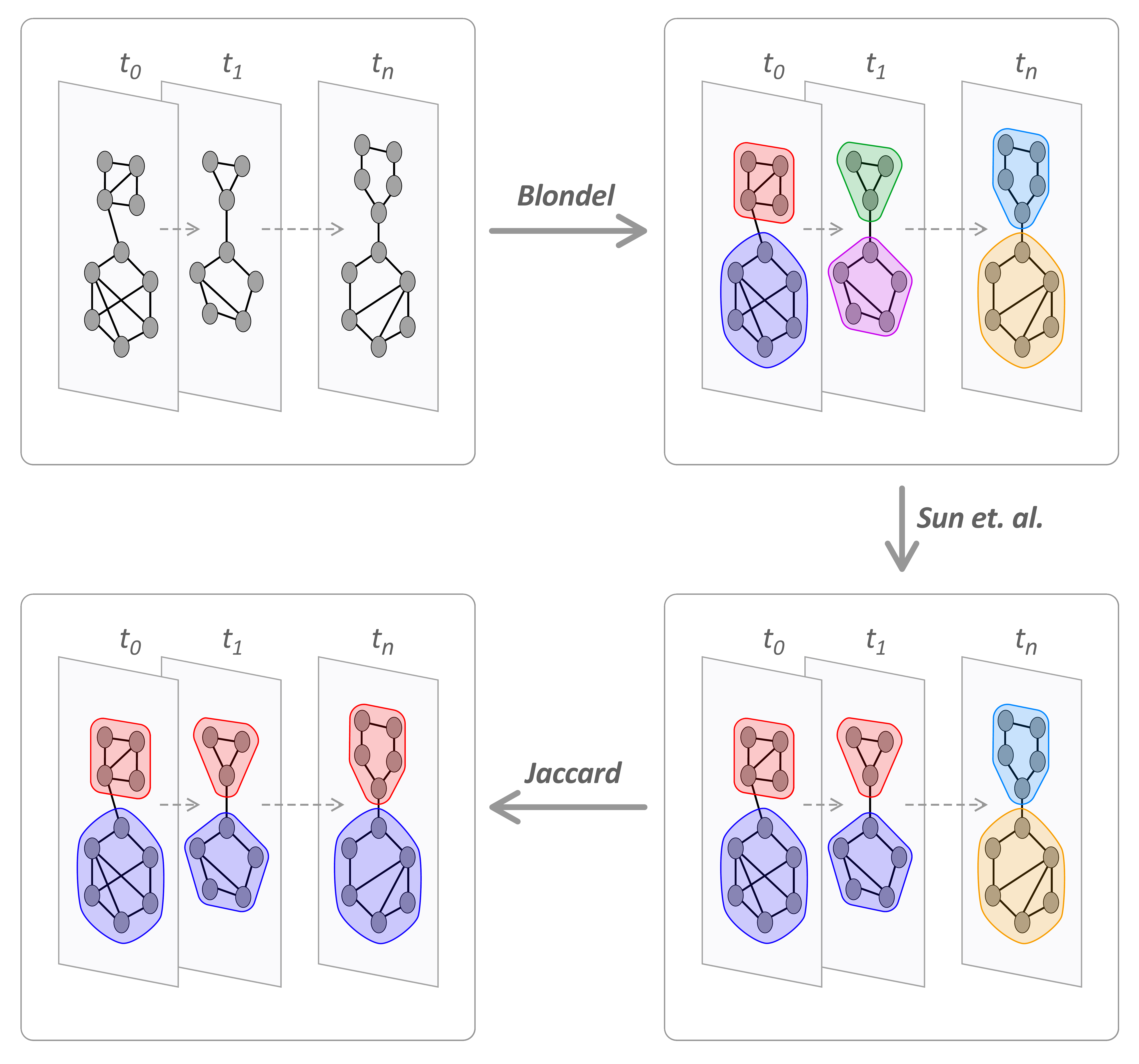}
	\caption{Proposed methodology for analyzing community dynamics in snapshot networks. Figure adapted from \cite{dakiche2019tracking} }
    \label{methodologyPipeline}
\end{figure}

Starting from the snapshots computed by \cite{vega2019spatio}, we followed the methodology illustrated in Figure \ref{methodologyPipeline}. Let the snapshots be a set of static graphs, $G = {G_0,G_1, ..., G_l}$, for each $G_i$ we apply the Louvain algorithm in a set of community structures $C^{i}={c_{0}^{i},...,c_{n}^{i}}$ where $n$ is the total of communities found in the snapshot $G_i$. It should be noted that each set $C^{i}$ is independent so that we should process such data to match communities that appeared in more than one snapshot. Thus, we process these structures regarding the approach defined by Sun et al. \cite{sun2015matrix}, which describes the rules to match communities from different snapshots according to the community life cycle. In other words, Sun et al. determine how to detect the birth of communities and the events that may occur in the life-cycle (split, merge, expansion, contraction) until its death. Nevertheless, we highlight that \cite{sun2015matrix} considers a consecutive sequence of snapshots, i.e., if a community $c^{i}_{j}$ was found in $G_i$ it should appear into the sequence $G_i=G_{i+1}$; if it could not find a community to match with $c^{i}_{j}$ in $G_{i+1}$, the authors consider that the community is dead. For example, if $c^{i}_{j}$ occurred again on $G_{i+2}$ or later, it will be considered as a new community birth with a new identification. 

To model fire dynamics and to describe the fire behavior in a given geographical area, we should detect communities that die and rebirth later, i.e., communities that resurges in a future snapshot. Such communities allow us to verify if fire events follow some periodicity over the years or not. Hence, to find resurgent communities we compare each community $c^{i}_j$ with all other communities into set $C^0, ..., C^{i-\Delta_t}$. We describe the community $c_{j}^{i}$ as a binary array, where positions indicate nodes, and the 1's determine the nodes that are part of the community. So, given such array, we may compare two communities, $c_{j}^{i}$ and $c_{k}^{i-\Delta_t}$, using the following Jaccard similarity equation:

\begin{equation}
  J(c_{j}^{i},c_{k}^{i-\Delta_t}) = \frac{M_{11}}{M_{01}+M_{10}+M_{11}},
 \label{eq_jaccardCoefInd}
\end{equation}

where $M{11}$ is the total number of positions where communities $c_{j}^{i}$ and $c_{k}^{i-\Delta_t}$ both have a value of 1, $M{01}$ represents the total number of position where the position of $c_{j}^{i}$ is 0 and the position of $c_{k}^{i-\Delta_t}$ is 1 and $M{10}$ the total number of position where the position of $c_{j}^{i}$ is 1 and the position of $c_{k}^{i-\Delta_t}$ is 0.

The construction process proposed in \cite{vega2019spatio} connects two consecutive wildfire events without any restriction about the geolocalization of the events. To consider the geolocalization into the model, we post-process the snapshots dataset $G$ to obtain the new set of snapshots, $G^{'} = \{G^{'}_{0},G^{'}_{1}, ..., G^{'}_{l}\}$, where two events in a snapshot $G_{i}$ are linked in $G_{i}^{'}$ if and only if the distance between the coordinates $(x,y)$ of the the cells representing the nodes $v_i$ and $v_j$ on grid is lower or equal than two, i.e., $d(v_i,v_j) = |(x_i-x_j)| + |(y_i-y_j)| \leq 2$. Thus, we submit $G^{'}$ in the same methodology illustrated by Figure \ref{methodologyPipeline}. 

\begin{figure}[h]
	\centering
	\begin{center}
		\subfigure[4-neighborhood]{\label{knn285}\includegraphics[scale=0.7]{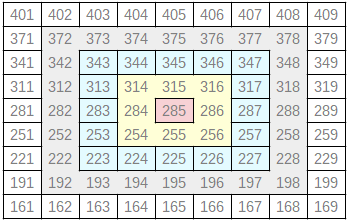}}\qquad
		\subfigure[Weighted distribution]{\label{knn285_pesos}\includegraphics[scale=0.7]{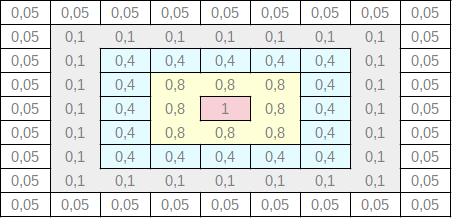}}
	\end{center}
	\caption{Example of neighborhood filtering and weights to the node 285: (a) we have the nodes until the 4th-cell-neighborhood, and (b) the weight distribution.}
	\label{fig_jaccardPond}
\end{figure}

We observe that the modified dataset provides communities with few nodes. We proceed using the geolocalization to merge communities that are near to each other on the grid, by adapting the resurgence process as follow: we describe a community $c_{j}^{i}$ as an array of real numbers, where each position on the array indicates a node of the network and its value indicate the weighted of the node into the community. For example, let $c^{i}_{j}$ be calculated to node 285 (Figure \ref{fig_jaccardPond}), then, the position 285 is set as 1, and the weight of its neighbors are computed from Equation \ref{eq_knn}:

\begin{equation}
  W_k=\frac{\kappa}{2^{k}},
 \label{eq_knn}
\end{equation}

\noindent where $\kappa$ is the parameter for defining the strength of the weighted distribution, and $k$ the radius of the nearest grid cells, i.e., neighborhood levels. In our analyses and without loss of generality, we adopted $\kappa = 0.8$. Figure \ref{fig_jaccardPond} illustrate the proposed filter: Figure \ref{knn285} shows node 285 and its k-neighborhood cells whereas Figure \ref{knn285_pesos} exemplifies the weights distribution. 

We always consider the higher value for a given node position, for example, if node 255 is part of community $c^{i}_{j}$, then the array position 255 will be set as 1, and their neighbors will be set as the maximum value between value on array and the value calculated by Equation \ref{eq_knn}. So, given such array we may compare two communities $c_{j}^{i} = {c_{j_1}^{i},...,c_{j_{900}}^{i}}$ and $c_{k}^{i-\Delta_t} = {c_{k_1}^{i-\Delta_t},...,c_{k_{900}}^{i-\Delta_t}}$, where 900 is the total number of cells and consequently the number of vertices, with Jaccard coefficient using the following equation:

\begin{equation}
  J_w(c_{j}^{i},c_{k}^{i-\Delta_t})=\frac{\sum_{x}^{900}min(c_{j_x}^{i},c_{k_1}^{i-\Delta_t})}{\sum_{x}^{900}max(c_{j_x}^{i},c_{k_1}^{i-\Delta_t})},
 \label{weigthedJaccard}
\end{equation}

The methodology described above provides a set of communities and their life cycle over the years. To answer our research questions (i), (ii) and (ii), we examined these communities on top of modularity values and events of resurgent, by using simple visual techniques. Details about these techniques and our experimental conclusions are presented in Section \ref{sec_results}. Additionally, we notice that modularity is a function that measures how good is the community division on a graph. According to Newman et al. \cite{newman2004finding}, values higher than 0.3 indicate that the graph was not brought forth by a random process, and therefore, we can claim that the community structure is relevant. Otherwise, although a community detection algorithm provides a graph partition, such a partition does not meet the concept of community division. So, before examining the communities provided by the Louvain algorithm, we first guarantee that these structures are valid from the theoretic perspective. 

\section{Results}
\label{sec_results}

\begin{figure}[htbp]
  \centering
	\includegraphics[scale=0.5]{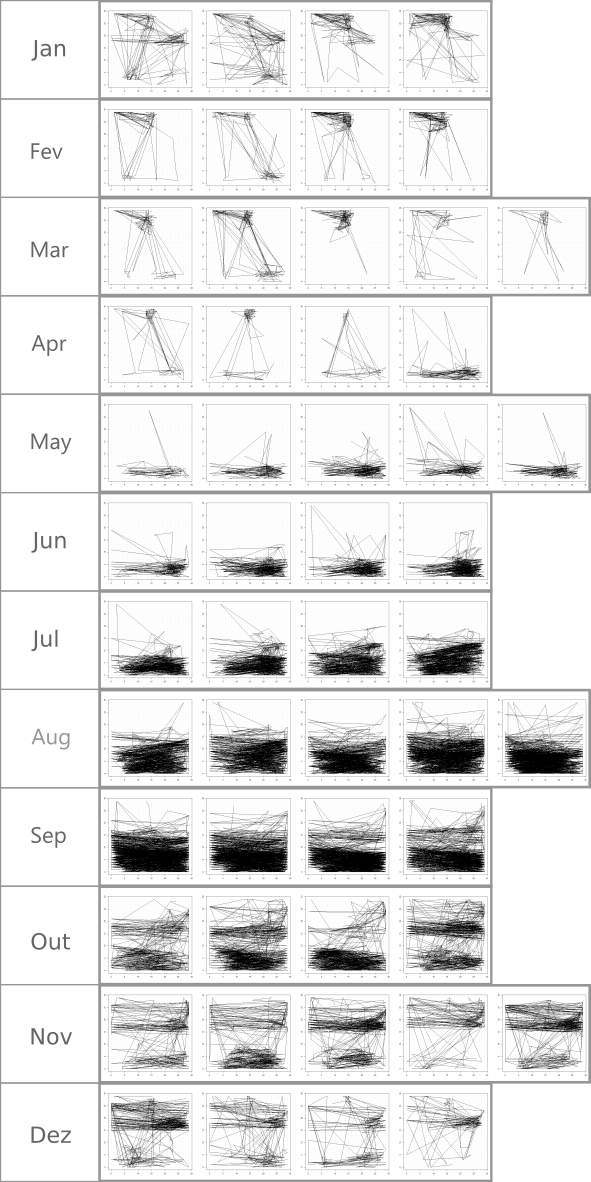}
	 \caption{The sequence of snapshots of the year 2003 organized by month. The nodes are located in a spatial layout that corresponds to the same grid cells disposition over the Amazon basin.}
	 \label{rede2003}
\end{figure}

A simple visual inspection reveals a pattern in the dynamics of wildfires in the Amazon Basin (Figure \ref{rede2003} illustrates this statement). By plotting the chronological network formed by data to the year 2003 over the $30 \times 30$ grid cells, we can observe that wildfires appear to occur more frequently in winter (June 21st) until mid-summer (mid of January). Furthermore, it is possible to note behavioral characteristics during this period. Starting the visual analysis from June snapshots, it can be noted that fire events seem to occur on the north of the grid, becoming more intense throughout July snapshots (i.e., the number of edges increase). Then, from August to September, they spread over most of the grid cells. After that, at the end of September to mid-summer, fire events split between south and north regions, forming two regions of fire activity in the snapshots, when finally, wildfires appear to lose their intensity. We notice that the fire events migrate to the south, starting a new cycle over the fall (March 20th to June 21st). It is noteworthy that this pattern tends to repeat in other snapshots from other years.

\begin{figure}[H]
	\centering
	\subfigure[Original chronnets snapshots]{\label{fig_modularityOli}\includegraphics[scale=0.60]{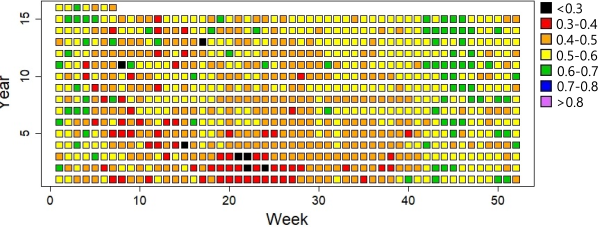}}
	\subfigure[2nd-cell-neighborhood Snapshots]{\label{fig_modularityOliRa}\includegraphics[scale=0.60]{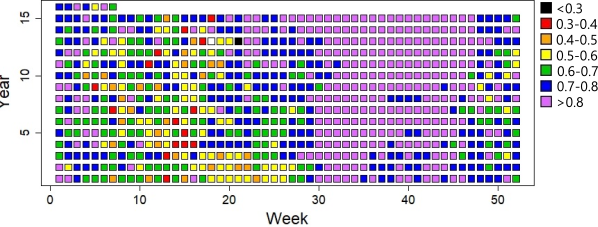}}
	\caption{Modularity value on each snapshot organized by week and years}
	\label{fig_modularity}
\end{figure}

\begin{figure}[!h]
  \centering
	\vspace{1mm}
	\includegraphics[scale=0.6]{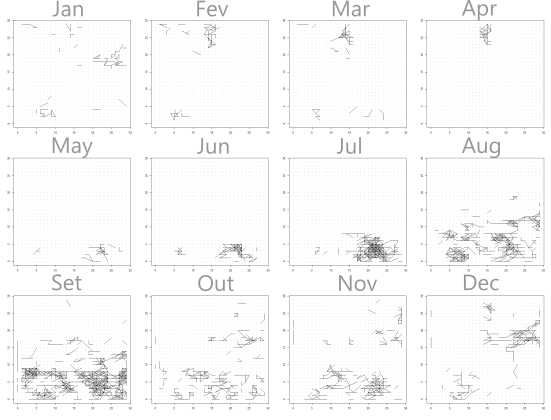}
	\caption{First week on each month of 2-neighborhood snapshots}
	\label{fig_redeOliveros_radius}
	\vspace{1mm}
\end{figure}

Here, we investigate if it is possible to model wildfires' behavior through communities, mainly from the communities which occur periodically over the years. Therefore, we verified the quality of all found communities. Figure~\ref{fig_modularity} shows the modularity value computed along the years analyzed. In the figure, each square represents a snapshot, and the color background points out the modularity value. Note that most snapshots show modularity value higher than 0.3, which is an indication of community structures in the analyzed data, according to Newman et al. \cite{newman2004finding}. We can also observe that filtering until the 2nd-cell-neighborhood in the snapshots present higher modularity values than the original Chronnets dataset. The before is due to the 2nd-cell-neighborhood construction provides snapshots formed by disconnected components (see, for instance, Figure \ref{fig_redeOliveros_radius}). Moreover, it is possible to notice a section of the year where modularity values are higher than other moments (please, observe the purple area in Figure \ref{fig_modularityOliRa} that is into the section where wildfires occur more frequently).

\begin{figure}[h]
  \centering
	\includegraphics[scale=0.6]{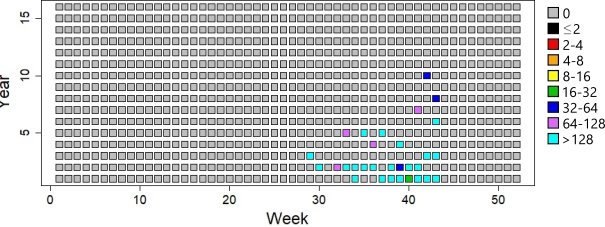}
	\caption{Example of dynamic behavior pattern detected in Chronnets snapshots. Colors indicate the size (number of nodes) of the communities.}
	\label{fig_patternOliveros}
\end{figure}

\begin{figure}[h]
	\centering
	\subfigure[No temporal pattern]{\label{fig_patternOliverosRad2}\includegraphics[scale=0.6]{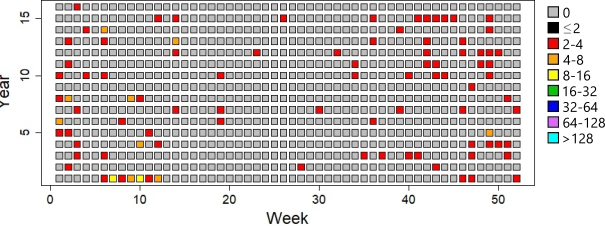}}
	\subfigure[Seasonal pattern]{\label{fig_patternOliverosRad1}\includegraphics[scale=0.6]{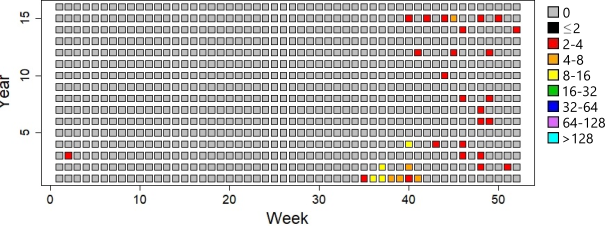}}
	\caption{Dynamic behavioral patterns detected in the 2nd-cell-neighborhood Snapshots. Colors indicate the size (number of nodes) of the communities.}
	\label{fig_patternOliverosRad}
\end{figure}

\begin{figure}[h]
	\centering
	\subfigure[No temporal pattern]{\label{fig_patternOliverosRadLac1}\includegraphics[scale=0.6]{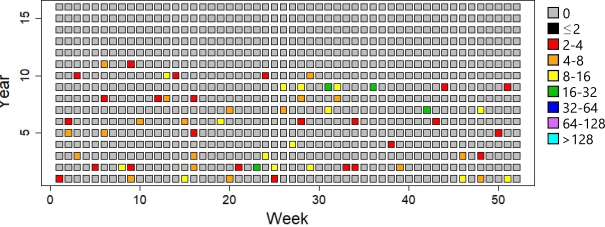}}
	\subfigure[Seasonal pattern]{\label{fig_patternOliverosRadLac2}\includegraphics[scale=0.6]{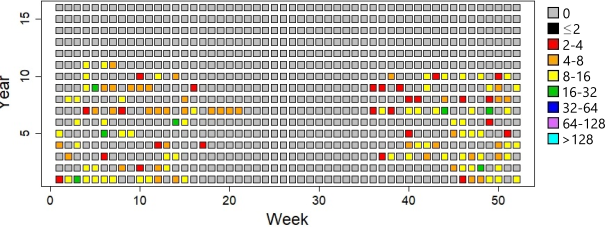}}
	\caption{Dynamic behavioral pattern detected after merging nearest communities in the 2nd-cell-neighborhood Snapshots. The color indicates the size (number of nodes) of the communities.}
	\label{fig_patternOliverosRadLac}
\end{figure}

The Figures \ref{fig_patternOliveros}, \ref{fig_patternOliverosRad}, and \ref{fig_patternOliverosRadLac} show the dynamic behavior pattern of the communities found on each set of experiments. In these Figures, a snapshot is illustrated as being a square where its background indicates if a given community occurs or not on this snapshot; gray color indicates the absence of the analyzed community, and other colors show the number of vertices that formed the community in the snapshot. 

Regarding the resurgence pattern of the community structures, our experiment suggests that Chronnets snapshots did not foment the detection of such structures. Using a threshold of 0.9, we could not found resurgent communities in Chronnets snapshots. A threshold of 0.9 means $90\%$ of vertices should be the same in both communities, so we relaxed it until 0.7 and found 2 and 12 structures, respectively (Figure~\ref{fig_patternOliveros}). On the other hand, filtering by the cell neighbors in the second radius, we found 221 resurgent communities setting the threshold to 0.9 (Figure~\ref{fig_patternOliverosRad}). However, we point out that few vertices form most of these communities. Even though the communities had few vertices, we observed that some of them were near each other. Therefore, we applied the weighted Jaccard coefficient  (Equation \ref{weigthedJaccard}) as an attempt to merge small communities which position on grid cells are close enough on grid cells. Through this approach, we can find 23 resurgent communities (Figure~\ref{fig_patternOliverosRadLac}). 

In this way, we found that the structures through original Chronnets snapshots did not revel all interesting patterns hold on the data set, even after relaxing the threshold. For instance, Figure \ref{fig_patternOliveros} shows the community found by using a threshold of 0.7. Through the 2nd-cell-neighborhood version (i.e., radius equals to two), we found interesting patterns using a threshold of 0.9 with the unweighted Jaccard approach (see Figure \ref{fig_patternOliverosRad}). When we assign weights in the 2nd-cell-neighborhood version, the communities detected by the Louvain algorithm + Sun et al. approach are prone to be joined as long as they are close in the vicinity of the grid cell, using the weighted Jaccard version (see Figure \ref{fig_patternOliverosRadLac}). Such patterns revel fire events that may occur at any time of the year (Figures \ref{fig_patternOliverosRad2} and \ref{fig_patternOliverosRadLac1}) and those who are expected year by year (Figures \ref{fig_patternOliverosRad1} and \ref{fig_patternOliverosRadLac2}). Finally, it is noteworthy to mention that all patterns seem to occur during the winter (June 21st) until mid-summer (mid of January). Besides, we could not find a relationship between the patterns and a specific geographical region.

\section{Conclusion and Future Works}
\label{sec_conFW}

Wildfires are environmental processes that have a significant impact on human life. Such phenomena can be described as a complex system that arouses scientific interest in researchers areas like global warming and climate change. Furthermore, wildfires are of practical interest for decision-makers like public agencies, ONGs, green peace, among others. 
In this context, network science has well-established tools to model complex systems. Although these tools might allow a better comprehension of the problem, to the best of our knowledge, just a few works have taken their advantages to model wildfires' dynamics. In the network science literature, the community structure is an excellent example of a concept that can support the study of fire dynamics, once this structure may further describe the network evolution. Here, understanding the evolution of the network means identifying where and when a fire event occurred, how they propagate and determining the regions and periods where such events are expected.

We have investigated here what kind of information the dynamic community structures could shed light on wildfires, especially regarding its temporal behavior. To achieve this goal, we explored a two-phase dynamic community detection approach, and, as proof of concept, we use the MODIS data set for fire events in the Amazon basin. Our results show that it is possible to model fire dynamics through community structures. The experiments indicate that the behavior of communities reveals fire events that can be detected at any time of the year and others that seem to occur at well-defined moments. However, we do not have sufficient evidence to relate a given geographical region for one of those kinds of events. Thus, our model can be improved through some knowledge discovery process. 

As future work, we can consider the following issues: (i) investigating grid structure with greater granularity, (ii) modeling wildfires as a source of streaming to preserve all temporal information and (iii) verifying if community detection methods developed to work with stream models may reveal new and relevant wildfire patterns. 

\section*{Acknowledgement}

This work was supported by the São Paulo Research Foundation (FAPESP), Grants 2015/50122-0, 2016/23642-6, 2016/23698-1, 2016/16291-2, 2017/05831-9, 2019/26283-5, and 2019/00157-3.

%
%
%

\bibliographystyle{splncs04Hacked}

%
\end{document}